\documentstyle[epsf,eqsecnum,aps,prd]{revtex}
\def\sqr#1#2{{\vcenter{\hrule height.#2pt
      \hbox{\vrule width.#2pt height#1pt \kern#1pt
          \vrule width.#2pt}
      \hrule height.#2pt}}}

\def\square{{\mathchoice{\sqr84}{\sqr84}{\sqr{5.0}3}{\sqr{3.5}3}}}
\def\Square{{}_s\!\mathop\square}
\def\edth{{\rlap{$\partial$}\raise0.3em\hbox{$-$}}}
\def\Si{{\rm Si}}
\def\Ci{{\rm Ci}}
\begin{document}
\hspace{-10mm}
\leftline{\epsfbox{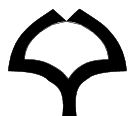}}
\vspace{-10.0mm} 
\thispagestyle{empty}
{\baselineskip-4pt
\font\yitp=cmmib10 scaled\magstep2
\font\elevenmib=cmmib10 scaled\magstep1  \skewchar\elevenmib='177
\leftline{\baselineskip20pt
\hspace{10mm} 
\vbox to0pt
   { {\yitp\hbox{Osaka \hspace{1.5mm} University} }
     {\large\sl\hbox{{Theoretical Astrophysics}} }\vss}}

\rightline{\large\baselineskip20pt\rm\vbox to20pt{
\baselineskip14pt
\hbox{OU-TAP-145}
\vspace{1mm}
\hbox{\today}\vss}}%
}
\vspace{8mm}
\begin{center}
{\large\bf
The Gravitational Reaction Force\\
on a Particle in the Schwarzschild Background
}\\
\smallskip

Hiroyuki Nakano\footnote{E-mail:denden@vega.ess.sci.osaka-u.ac.jp}
and
Misao Sasaki\footnote{E-mail:misao@vega.ess.sci.osaka-u.ac.jp}
\\
\smallskip

{\em Department of Earth and Space Science,~Graduate School of
  Science Osaka University,\\ Toyonaka, 560-0043, Japan
}\\
\smallskip

\end{center}


\begin{abstract}
We formulate a new method to calculate the gravitational
reaction force on a particle of mass $\mu$ orbiting a massive
black hole of mass $M$. In this formalism, the tail part of
the retarded Green function, which is responsible for the
reaction force, is calculated at the level of
the Teukolsky equation. Our method naturally allows a systematic
post-Minkowskian (PM) expansion of the tail part at short
distances. As a first step, we consider the case of
a Schwarzschild black hole and explicitly
calculate the first post-Newtonian (1PN) tail part of the Green
function. There are, however, a couple of issues to be resolved before
explicitly evaluating the reaction force by
applying the present method. We discuss possible resolutions of these
issues.

\end{abstract}


\section{Introduction}
In the black hole perturbation approach to gravitational
radiation from an inspiralling compact binary, one calculates
the gravitational waves emitted by a point particle orbiting
a black hole, assuming the mass of the particle ($\mu$) is
much less than the mass of the black hole ($M$).
At the lowest order in $\mu/M$, the particle moves along
a geodesic trajectory in the black hole geometry.
Already at this lowest order, this approach has been
proved to be very powerful for evaluating general
relativistic corrections to the gravitational wave forms
from a compact binary, even for neutron star-neutron star
(NS-NS) binaries\cite{MSSTT}.
However, in such calculations one inevitably has to assume
that the radiation reaction time scale is sufficiently
long compared to the orbital period, so that the
adiabatic approximation to the orbital evolution is valid.
That is, the orbital evolution is determined by
the energy and angular momentum balance between the orbit
and the graviational radiation.

However, there are several situations in which
such an approximation cannot be justified.
For example, when the particle is
nearly in the last stable circular orbit, the
radiation reaction time scale becomes comparable
to the orbital period. When the particle is in a very
eccentric orbit or in a scattering orbit, presumably
one has to determine the orbital evolution not by
averaging over many periods but by each
scattering event.

Another case is when the massive object is in fact a rotating
black hole. In this case the geometry is Kerr, and the geodesic
orbit is determined not only
by the energy and the $z$-component of the angular momentum
but also by the Carter constant\cite{Carter}. Since the Carter constant
is not associated with the Killing vector field of the
Kerr geometry, its evolution cannot be determined
by evaluating the gravitational radiation at infinity, 
even though the adiabatic approximation may still be valid.

In order to treat these situations, it is therefore necessary
to evaluate the correction to the
equation of motion of a particle in the background geometry.
That is, we must know the reaction force of $O(\mu/M)$
acting on the particle,
\begin{eqnarray}
{D^2z^\mu(\tau)\over d\tau^2}=F^\mu(z)\,.
\label{eq:for}
\end{eqnarray}
Formally $F^\mu$ is given by \cite{MiSaTa,QuiWal}
\begin{eqnarray}
F^\mu=-{1\over2}(g^{\mu\nu}+u^\mu u^\nu)
\left(
2h^{\rm tail}_{\nu\beta;\alpha}-h^{\rm tail}_{\alpha\beta;\nu}
\right)u^\alpha u^\beta\,,
\label{eq:force}
\end{eqnarray}
where $u^\mu=dz^\mu/d\tau$ and $h^{\rm tail}_{\mu\nu}$ is the
tail part of the linear perturbation caused by the particle,
i.e., the contribution from a part of the retarded Green function
that has support inside the past lightcone.
Thus our task is to evaluate the tail part of the
Green function\footnote{In this paper, 
we refer to the `retarded Green function' 
simply as the `Green function'.}.

However, since the tail part depends non-locally on the
background geometry, it seems practically impossible to
evaluate it directly.
As a way to avoid this difficulty, Mino proposed a
very interesting approach\cite{Mino}. He noticed that the direct
part of the Green function, which has support only on the
lightcone, is determined by the local geometry. 
Hence the
corresponding metric perturbation can be
evaluated solely with knowledge of the local geometry.
On the other hand, in the black hole background, the
linear perturbation is known to be described by
the Teukolsky formalism\cite{Teuk}, 
 and there exists a method to obtain
the general solution in a systematic manner\cite{MaSuTa}.
Then, using the transformation obtained by Chrzanowski\cite{Chrz},
one can construct the corresponding metric perturbation.
Hence, on the one hand, one may construct the full
metric perturbation by solving the Teukolsky equation,
and on the other hand, one can calculate the direct part
of the metric perturbation by analyzing the local geometry.
One can then obtain the tail part by subtracting the
direct part from the full metric perturbation.

This approach was further advanced by Mino and
Nakano\cite{MinNak}, and they succeeded in obtaining some interesting
results. Yet, their method seems rather cumbersome.
It involves many steps of lengthy calculations, mainly
resulting from the different choice of gauges in which the full
metric perturbation and its direct part are evaluated.

In this paper, we propose a new, possibly more practical method to
evaluate the tail part of the perturbation.
As noted by Mino \cite{Mino}, what one needs to do is to extract 
the direct part of the Green function.
But this does not have to be done on the level of the
metric perturbation. Instead, if one considers
the Teukolsky equation in the spacetime domain,
one can evaluate the direct part of the Green function
for the Teukolsky equation directly, without transforming
it to the metric perturbation. By Fourier-harmonic expanding
the direct part obtained in this manner, one can then systematically
evaluate the tail part. Then the corresponding
metric perturbation and the reaction force $F^\mu$ can be
calculated by using the Chrzanowski transformation\cite{Chrz}.

A technical problem in this approach is that the Teukolsky
equation has a very non-trivial angular spin weight dependence,
even in the limit of flat spacetime, $M\to0$.
Thus the evaluation of the direct part 
becomes very involved.
As a way to avoid this problem, we consider the transformation of the
Teukolsky equation with spin weight $s$ to a Klein-Gordon-type
equation (i.e., with spin weight $s=0$). In the case of the
Schwarzschild background, this is essentially
the Chandrasekhar transformation that transforms the Teukolsky
equation into the Regge-Wheeler equation\cite{Chantrans}. In the case of
the Kerr background, there is no known transformation
of this kind. One possibility is to use the
transformation given by Sasaki and Nakamura \cite{SasNak}
in a perturbative manner, with respect to the spin parameter
$a/M$ of the Kerr geometry. In this paper, however,
we focus on the case of the Schwarzschild black hole
background and defer the extension
to the Kerr case to a future work.

There are also a couple of issues to be resolved before applying the
present method to the actual calculation of the reaction force.
First, the Teukolsky equation, by its nature, does not give the
$\ell=0$ and 1 spherical harmonic components of the perturbation. Hence
the tail part of these components must be calculated separately.
Second, and most importantly, since the reaction force $F^\mu$ is a
gauge-dependent notion, its physical effect can be clarified only after
we obtain the $O(\mu)$ correction to the gravitational radiation emitted
to infinity, but what we have at hand is only the linear-order metric
perturbation. Thus, strictly speaking, there is no point in evaluating
the reaction force unless we have a method to evaluate the
corresponding correction to the gravitational radiation.

There have been other proposals for the regularization of the self
force. Barack and Ori\cite{BarOri} proposed a mode-sum regularization
scheme. The calculation of the reaction force in their approach has
two steps. 
First, each Fourier-harmonic mode of the bare self force is
calculated.
Then the sum over all modes is made subject to certain 
regularization procedure, which requires several regularization 
parameters.
The calculation of the bare force on a scalar charge is done
in \cite{Bur1,Bur2,BarBur}. 
The analytical determination of the regularization
parameters is carried out using a local perturbative 
analysis in \cite{BarOri,Bar}. 
A method similar to that in \cite{BarOri} was proposed independently
by Lousto\cite{Lousto}.
Very recently, using the mode-sum regularization
prescription, the regularization of the scalar and electromagnetic
self force in the spacetime of spherical shells has been discussed
by Burko, Liu and Soen\cite{BuLiSo}.

This paper is organized as follows.
In Sec.~\ref{sec:RWGF}, we consider the Chandrasekhar transformation
in the spacetime domain and discuss the nature of the direct part and
the tail part of the Regge-Wheeler Green function.
Then we give the Fourier-harmonic expanded form of the Teukolsky
and Regge-Wheeler Green functions.
In Sec.~\ref{sec:1PMtail},
we formulate a method to calculate the tail part of the Regge-Wheeler
 Green function. Then, the tail part correct to the first
 post-Minkowskian (1PM) order is given under the short distance
 approximation, and an explicit expression of the tail in the
first post-Newtonian (1PN) order is given.
Then in Sec.~\ref{sec:RV}, the procedure to obtain the regularized
reaction force is discussed.
Finally, in Sec.~\ref{sec:discussion}, we summarize the results and
discuss a possible way to resolve the remaining issues mentioned above.


\section{Regge-Wheeler and Teukolsky Green Functions}\label{sec:RWGF}

\subsection{Chandrasekhar transformation in
spacetime domain}\label{subsec:Chandra}

The Teukolsky equation in the spacetime domain is expressed as
\begin{eqnarray}
\Square\,{}_s\!\Psi\equiv
\left[\nabla^\mu\nabla_\mu+{2s\over r^2}
\left\{\left(-r+{Mr^2\over\Delta_K}\right){\partial_t}
+(r-M){\partial_r}+i{\cos\theta\over\sin^2\theta}\partial_\varphi
+{1-s\cot^2\theta\over2}\right\}\right]{}_s\!\Psi=-4\pi\,{}_s{\cal T}\,,
\label{teuk}
\end{eqnarray}
where $s=0$, $\pm1$ or $\pm2$, and 
$\Delta_K=r(r-2M)$. We note that ${}_s\!\Psi$ has 
spin weight $s$ with respect to the spatial rotation of
the tetrad, but it is a spacetime scalar. 
The quantity ${}_s{\cal T}$ is the source
term whose explicit form can be found in \cite{Teuk}.
Although our interest is in the gravitational perturbation,
and it is described by the Teukolsky equation with $|s|=2$,
we leave the value of $s$ unfixed until we need to fix it.
As we see below, this makes it easy to identify
the tail part of the Green function.

The difficulty of directly dealing with the Teukolsky equation
comes from the non-trivial angular dependence
in its zeroth and first derivative terms. In particular,
this causes non-conventional behavior of the Green function
even in the flat spacetime limit.
To overcome this difficulty, we consider a transformation
of the Teukolsky equation. Since there is a one-to-one
correspondence between the Teukolsky equation
with spin weights $s$ and $-s$ (the Starobinsky-Teukolsky identities),
we concentrate on negative spin weights 
$s=0$, $-1$ and $-2$.
For $s\le0$, we introduce a new field variable ${}_sX$ by setting
\begin{eqnarray}
{}_s\!\Psi=\bar\edth^{|s|}
\left({\Delta_K \over r}\right)^{|s|}
\left(\partial_r-{r^2\over\Delta_K}\partial_t\right)^{|s|}r^{|s|}\,{}_sX\,,
\label{chantrans}
\end{eqnarray}
where
\begin{eqnarray}
\edth&=&-\left(\partial_\theta+{i\over\sin\theta}\partial_\varphi
-s\cot\theta\right),
\nonumber\\
\bar\edth&=&-\left(\partial_\theta-{i\over\sin\theta}
\partial_\varphi+s\cot\theta\right).
\label{edthdef}
\end{eqnarray}
It should be noted that $\edth$ ( $\bar\edth$ ) operating 
on a quantity of spin weight $s$ yields a quantity of
spin weight $s+1$ ($s-1$). In particular, we have
\begin{eqnarray}
\edth\,({}_sY_{\ell m})&=&\left[\ell(\ell+1)-s(s+1)\right]^{1/2}
{}_{s+1}Y_{\ell m}\,,
\nonumber\\
\bar\edth\,({}_sY_{\ell m})&=&-\left[\ell(\ell+1)-s(s-1)\right]^{1/2}
{}_{s-1}Y_{\ell m}\,,
\nonumber\\
\bar\edth\edth\,({}_sY_{\ell m})&=&-\left[\ell(\ell+1)-s(s+1)\right]
{}_sY_{\ell m}\,,
\label{eq:Ed}
\end{eqnarray}
where the ${}_sY_{\ell m}$ are the normalized
spherical harmonics of spin weight $s$. This implies 
that ${}_sX$ has spin weight $s+|s|=0$ for $s\le0$.
If expanded in terms of $e^{-i\omega t}$ and ${}_sY_{\ell m}$,
the radial part of the above transformation is just the
Chandrasekhar transformation that relates homogeneous Teukolsky
functions to Regge-Wheeler functions, apart from the overall
normalization constant that depends on $\ell$ and $\omega$. A
conceptually important difference is that here we do not associate
${}_sX$ with the spin weight $s$, but with $s=0$.

Inserting (\ref{chantrans}) into (\ref{teuk}), we find
\begin{eqnarray}
\Square\,{}_s\!\Psi
&=&\bar\edth^{|s|}\left({\Delta_K \over r}\right)^{|s|}{1\over r^2}
\left(\partial_r-{r^2\over\Delta_K}\partial_t\right)^{|s|}r^{|s|+2}
\left[\nabla^\mu\nabla_\mu+s^2\,{2M\over r^3}\right]\,{}_sX=
-4\pi\,{}_s{\cal T}\,,
\nonumber\\
&&\Rightarrow
\quad
\left[\nabla^\mu\nabla_\mu+s^2\,{2M\over r^3}\right]\,{}_sX=
-4\pi\,{}_s{\tilde {\cal T}}\,.
\label{Xeq}
\end{eqnarray}
We see that the equation for ${}_sX$ has the standard
Klein-Gordon type structure with a potential term
that vanishes sufficiently rapidly at infinity irrespective
of the spin weight. Note that its radial part is just the
Regge-Wheeler equation. It then follows that ${}_sX$ can be naturally
expanded in terms of the scalar (i.e., $s=0$) spherical harmonics even
for nonzero $s$, in accordance with the discussion above.
A very important point to be noted is the following.
If one constructs the Green function of Eq.~(\ref{Xeq}) for
${}_sX$, its direct part is exactly the same for all
the spin weights. This is because the local causal structure of
a hyperbolic equation depends only on the derivative
terms of the equation (i.e., the $\nabla^\mu\nabla_\mu$ part).
This implies that the part that should be discarded from
the Green function is independent of $s$ if done for
the variable ${}_sX$. In other words, any
$s$-dependent terms in the Green function for ${}_sX$
belong to the tail part.


\subsection{Green functions for ${}_s\Psi$ and ${}_sX$}\label{subsec:GF}

We first consider the Green function ${}_sG^{\rm ret}$ for
Eq.~(\ref{Xeq}). It satisfies
\begin{eqnarray}
  \label{Greteq}
\left[\nabla^\mu\nabla_\mu+s^2\,{2M\over r^3}\right]\,
{}_sG^{\rm ret}(x,x')
=-{\delta(t-t')\delta(r-r')\delta^2(\Omega-\Omega')
\over rr'}\,.
\end{eqnarray}
Its Fourier-harmonic expansion is expressed as
\begin{eqnarray}
  \label{Gret}
 {}_sG^{\rm ret}(x,x')&&=\int{d\omega\over2\pi}\sum_{\ell m}
\,{}_sG_{\ell m\omega}(r,r')\,
{}_0Y_{\ell m}(\Omega)\,\overline{{}_0Y_{\ell m}(\Omega')}\,
e^{-i\omega(t-t')} \,;
\nonumber\\
&&{}_sG_{\ell m\omega}(r,r')=-{\omega\over|\omega|}
{1 \over W(X_\ell^{\rm up},X_\ell^{\rm in})} \,
{X_\ell^{\rm up}(r_{>})\,X_\ell^{\rm in}(r_{<})
\over rr'}\,,
\end{eqnarray}
where $r_{>}={\rm max}\{r,r'\}$, $r_{<}={\rm min}\{r,r'\}$,
${}_0 Y_{\ell m}$ is the usual (spin weight $0$) spherical harmonic
function,
$W(X_\ell^{\rm up},X_\ell^{\rm in})$ is the Wronskian,
\begin{eqnarray}
  \label{wronskian}
  W(X_\ell^{\rm up},X_\ell^{\rm in})
=\left({d\over dr^*}X_\ell^{\rm up}\right)X_\ell^{\rm in}
-X_\ell^{\rm up}{d\over dr^*}X_\ell^{\rm in}\,,
\end{eqnarray}
and $X_\ell^{\rm up}$ and $X_\ell^{\rm in}$ are the
homogeneous solutions of the Regge-Wheeler equation
having the asymptotic behaviors,\footnote{In this paper, we follow
the notation of Chrzanowski and Misner\cite{ChrMis} for the
homogeneous solutions.}
\begin{eqnarray}
  X_\ell^{\rm in}&\sim&\left\{
    \begin{array}{ll}
    {\cal A}_{\ell\omega}^{\rm in}e^{-i\omega r^*}
         +{\cal A}_{\ell\omega}^{\rm out}e^{i\omega r^*}&~(r\to\infty),\\
     e^{-i\omega r^*}&~(r\to2M),
    \end{array}\right.
\nonumber\\
  X_\ell^{\rm up}&\sim&\left\{
    \begin{array}{ll}
      e^{i\omega r^*}&~(r\to\infty),\\
     {\cal D}_{\ell\omega}^{\rm in}e^{-i\omega r^*}
         +{\cal D}_{\ell\omega}^{\rm out}e^{i\omega r^*}&~(r\to2M).
    \end{array}\right.
\end{eqnarray}
It then follows that $W=2i\omega{\cal A}_{\ell\omega}^{\rm in}
=2i\omega{\cal D}_{\ell\omega}^{\rm out}$.

To 1PM order, the necessary formulas for
the radial functions $X_\ell^{\rm in}$ and $X_\ell^{\rm up}$
may be found in Poisson and Sasaki\cite{PoiSas} and Leonard and
Poisson\cite{LeoPoi}. (Their explicit expressions are given in
Appendix \ref{1PMX}.) For $\omega>0$, they take the form
\begin{eqnarray}
X_\ell^{\rm in}(z)
&&=z\left[j_\ell(z)+\epsilon\,\xi_\ell^{\rm in}(z)
+O(\epsilon^2)\right],
\label{Xin}\\
X_\ell^{\rm up}(z)
&&=z\left[h^{(1)}_\ell(z)+\epsilon\,\xi_\ell^{\rm up}(z)
+O(\epsilon^2)\right],
\label{Xup}
\end{eqnarray}
where $\epsilon=2M\omega$, $z=\omega r$, and the functions
$\xi_\ell^{\rm in}$ and $\xi_\ell^{\rm up}$
are defined in Eqs.~(\ref{1PMXin}) and (\ref{1PMXup}).
As noted in Appendix \ref{1PMX}, for $\omega<0$
(i.e, for $z<0$), they should be replaced by
$\overline{X^{\rm in}_\ell(|z|)}$ and $\overline{X^{\rm up}_\ell(|z|)}$,
respectively.
With this understanding, we find
\begin{eqnarray}
{}_sG_{\ell m\omega}(r,r')=i\omega\,
\left[j_\ell(z_{<})h^{(1)}_\ell(z_{>})
+\epsilon\,\left(j_\ell(z_{<})\xi^{\rm up}_\ell(z_{>})
                +\xi^{\rm in}_\ell(z_{<})h^{(1)}_\ell(z_{>})\right)
+O(\epsilon^2)\right].
\label{1PMG}
\end{eqnarray}
Hence
\begin{eqnarray}
\sum_{m}{}_sG_{\ell m\omega}
&&Y_{\ell m}(\Omega)\overline{Y_{\ell m}(\Omega')}
\nonumber\\
&&={i\omega\over4\pi}\,(2\ell+1)
\left[j_\ell(z_{<})h^{(1)}_\ell(z_{>})
+\epsilon\,\left(j_\ell(z_{<})\xi^{\rm up}_\ell(z_{>})
                +\xi^{\rm in}_\ell(z_{<})h^{(1)}_\ell(z_{>})\right)
+O(\epsilon^2)\right]P_{\ell}(\mu)\,,
\end{eqnarray}
where $\mu=\bbox{\Omega}\cdot\bbox{\Omega}'$.

Now, given the Green function for ${}_sX$, the Green function for
${}_s\Psi$ is given as follows.
The Teukolsky functions ${}_sR_{\ell}^{\rm in}$,
${}_sR_{\ell}^{\rm up}$ and
the Regge-Wheeler functions ${}_sX_{\ell}^{\rm in}$,
${}_sX_{\ell}^{\rm up}$
are related by the Chandrasekhar transformation,
\begin{eqnarray}
{}_sR_{\ell}^{\rm in}
= {}_s\chi_{\ell}^{\rm in}\,{}_sC_{\omega}\,{}_sX_{\ell}^{\rm in} \,,
\quad
{}_sR_{\ell}^{\rm up}
={}_s\chi_{\ell}^{\rm up}\,{}_sC_{\omega}\,{}_sX_{\ell}^{\rm up} \,,
\label{eq:CX}
\end{eqnarray}
where ${}_sC_\omega$ ($s=0,\pm1,\pm2$) are the Chandrasekhar operators
\cite{Chantrans,PoiSas,LeoPoi}
and ${}_s\chi_{\ell}^{\rm in}$ and ${}_s\chi_{\ell}^{\rm up}$ are
certain normalization coefficients (the case of $s=0$ is trivial,
since ${}_0R={}_0X$).
For $s=-2$, the operator ${}_{-2}C_{\omega}$ is defined by
\begin{eqnarray}
C_{\omega} = \omega r^2f\,{\cal L}\,f^{-1}{\cal L}\,r \,,
\label{Chop}
\end{eqnarray}
where
\begin{eqnarray}
{\cal L} = f{d \over dr} +i\,\omega \,,
\quad f=\left(1-{2M\over r}\right),
\label{eq:L}
\end{eqnarray}
and we have omitted the spin index of ${}_{-2}C_\omega$
for simplicity and denoted it simply by $C_\omega$, 
since we are interested only in the case $s=-2$.
The coefficients ${}_{-2}\chi_{\ell}^{\rm in}$ and
${}_{-2}\chi_{\ell}^{\rm up}$ are given by
\begin{eqnarray}
\chi_{\ell}^{\rm in}
 = {16\,(1-2\,i\,M\omega)(1-4\,i\,M\omega)(M\omega)^3
\over (\ell-1)\ell(\ell+1)(\ell+2)-12\,i\,M\omega} \,,
\quad
\chi_{\ell}^{\rm up} &=& -{1 \over 4} \,,
\label{chielldef}
\end{eqnarray}
where we have also omitted the spin indices for simplicity.
In addition, in the rest of the paper, we denote ${}_{-2}\Psi$ by
$\psi_4$, in accordance with the conventional notation for the
Weyl scalar.

The Green function for the Teukolsky equation is constructed from the
above homogeneous functions as given by Eq.~(\ref{Gteuk}) below.
Except for the Wronskian $W(R_\ell^{\rm up},R_\ell^{\rm in})$,
however, the explicit form of the Green function for the Teukolsky
equation is not needed for the
following calculation, as will be clarified later.


\section{Tail Part at Short Distances}\label{sec:1PMtail}

As usual, we write the Green function
in the Hadamard form,
\begin{eqnarray}
{}_sG^{\rm ret}(x,x')=\theta(\Sigma(x),x')\left(
{u \over4\pi}\delta(\sigma)+\theta(-\sigma)\,{}_sv\right),
\label{Hadamard}
\end{eqnarray}
where $\sigma$ is half the squared geodetic distance between $x$ and $x'$,
$u$ is a bi-scalar that depends only on the local geometry,
and ${}_sv$ is a bi-scalar that describes the tail effects.
We have a unique solution for each of these bi-scalars.
As noted in the previous section, the direct part of
 ${}_sG^{\rm ret}$ is independent of $s$, and the bi-scalar $u$
is given simply by that in the case of the scalar
d'Alembertian\cite{DeWBre},
\begin{eqnarray}
  u=\left({-\det|\sigma_{;\alpha;\beta'}|
\over\sqrt{g(x)g(x')}}\right)^{1/2}
\end{eqnarray}
Let us denote the direct part of ${}_sG^{\rm ret}$ by
$G^{\rm dir}$:
\begin{eqnarray}
  \label{Gdirect}
  G^{\rm dir}(x,x')=\theta(\Sigma(x),x')\,{u\over4\pi}\,\delta(\sigma).
\end{eqnarray}
Our task is to evaluate $G^{\rm dir}$ in the Fourier-harmonic expanded
form at short distances.

Since the full evaluation of the direct part is difficult,
we consider its post-Minkowskian (PM) expansion.
To do so, it is more convenient to work in the isotropic coordinates
than in the Schwarzschild coordinates.
The radial coordinates of these two coordinates
are related as $r=r_I(1+M/2r_I)^2$.
Here $r$ denotes the conventional Schwarzschild radial coordinate
and $r_I$ denotes the isotropic radial coordinate.
It should be noted that the isotropic coordinates satisfy
the harmonic gauge condition to 1PM order.
The bi-scalar $\sigma(x,x')$ accurate to 1PM order
can be found in a classic paper by Thorne and Kovacs\cite{ThoKov},
or in Leonard and Poisson\cite{LeoPoi}.

We will come back to the explicit evaluation of the 1PM tail part
later. Here, we first describe the formal procedure to obtain the
tail part.
Let us express the Fourier-harmonic components of the full
Regge-Wheeler Green function (\ref{Gret}) as
\begin{eqnarray}
{}_sG(\ell m\omega;\, x,x')
={}_sG_{\ell m\omega}(r,r') \,
{}_0 Y_{\ell m}(\Omega)\overline{{}_0 Y_{\ell m}(\Omega')}\,
e^{-i\omega(t-t')} \,.
\end{eqnarray}
Correspondingly, the Fourier-harmonic components of the
direct part, which is $s$-independent, may be expressed as
\begin{eqnarray}
G^{\rm dir}(\ell m\omega;\,x,x')
 = G_{\ell m\omega}^{\rm dir}(r,r')
\,{}_0 Y_{\ell m}(\Omega) \overline{{}_0 Y_{\ell m}(\Omega')}\,
e^{-i\omega(t-t')} \,.
\end{eqnarray}
The tail part of the Green function, i.e., the regularized Green
function, is obtained by subtracting the direct part from
the full Green function: 
\begin{eqnarray}
{}_s G^{\rm tail}(\ell m\omega;\,x,x')
=&&{}_s G^{\rm tail}_{\ell m\omega}(r,r')
\,{}_0 Y_{\ell m}(\Omega)\overline{{}_0 Y_{\ell m}(\Omega')}\,
e^{-i\omega(t-t')} \,;
\nonumber\\
{}_s G^{\rm tail}_{\ell m \omega}(r,r')
&&= {}_s G_{\ell m\omega}(r,r')
- G^{\rm dir}_{\ell m\omega}(r,r') \,.
\end{eqnarray}
Since this subtraction is done on a ``mode-by-mode'' basis,
we call our procedure the ``mode-by-mode regularization'' of
the Green function.

Now let us return to the evaluation of the first post-Minkowskian
(1PM) order tail. To 1PM order,
the geodesic connecting $x$ and $x'$ can be approximated
by a straight line on the flat background\cite{ThoKov,LeoPoi}:
\begin{eqnarray}
\xi^\mu(\lambda)=x'{}^\mu+\lambda X^\mu\,;
\quad X^\mu=x^\mu-x'{}^\mu\,.
\end{eqnarray}
Here the affine parameter $\lambda$ is normalized so that
$\xi^\mu(0)=x'{}^\mu$ and $\xi^\mu(1)=x^\mu$.
We then have
\begin{eqnarray}
\sigma(x,x')&=&{1\over2}
\left(\eta_{\mu\nu}
+\int_0^1 h_{\mu\nu}(\xi(\lambda))d\lambda\right)X^\mu X^\nu
\nonumber\\
&=&{1\over2}\left(-\Delta t^2(1-F)+R^2(1+F)\right)\,;
\qquad F=\int_0^1{2M\over r_I(\lambda)}d\lambda\,,
\end{eqnarray}
where $\Delta t=t-t'$ and
$R^2=\delta_{ij}(x^i-x'{}^i)(x^j-x'{}^j)$.
The 1PM correction factor $F$ can be calculated exactly as 
\begin{eqnarray}
F&=&{2M\over R}\ln{(r_I+r'_I+R)\over(r_I+r'_I-R)}
={2M\over r_I+r'_I}\sum_{n=0}^\infty
{2\over2n+1}\left({R\over r_I+r'_I}\right)^{2n}
\nonumber\\
&=&{4M\over r_I+r'_I}+{4M\over 3(r_I+r'_I)}
\left({R\over r_I+r'_I}\right)^{2}
+O(R^4),
\end{eqnarray}
where the short-distance approximation is applied to the
second line.
The bi-scalar $u$ accurate to 1PM order is unity, 
because of the Ricci flatness of the Schwarzschild geometry.
Thus the direct part of the Green function becomes
\begin{eqnarray}
 G^{\rm dir}(x,x')&=&{\theta(\Delta t)\over4\pi}\,2
\delta\left(-\Delta t^2(1-F)+R^2(1+F)\right)
\nonumber\\
&=&{\delta\left(\Delta t-R(1+F)\right)\over 4\pi R}\,.
\end{eqnarray}

To subtract the direct part from the Green function,
we take the Fourier-harmonic expansion of the direct part.
The Fourier transformation of $G^{\rm dir}$ is
\begin{eqnarray}
\tilde G^{\rm dir}_\omega(\bbox{x},\bbox{x'})
&=&\int G^{\rm dir}e^{i\omega \Delta t}d\Delta t
={e^{i\omega R(1+F)}\over 4\pi R}
\nonumber\\
&=&{e^{i\omega R}\over4\pi R}\left(1+i\omega RF+O(M^2)\right)\,.
\end{eqnarray}
Using the relation 
\begin{eqnarray}
{e^{i\omega R}\over R} = \sum_{\ell=0}^\infty
 (2\,\ell+1)\,j_{\ell}(\omega r_{I<})\,
h^{(1)}_{\ell}(\omega r_{I>})\,P_{\ell}(\mu) \,,
\end{eqnarray}
and re-expressing $r_I$ in terms of the Schwarzschild radial coordinate
$r$, we find
\begin{eqnarray}
\tilde G^{\rm dir}_\omega(\bbox{x},\bbox{x'})
&&= \sum_{\ell=0}^\infty
{i\omega\over4\pi}\,(2\ell+1)
\Biggl[j_\ell(z_{<})h^{(1)}_\ell(z_{>})
+{\epsilon\over2}\,\Biggl(
\left({1\over z_{<}}+{1\over z_{>}}\right)
j_\ell(z_{<})\,h^{(1)}_\ell(z_{>})\nonumber \\
&& \quad \quad \quad \quad
+{z_{<} \over z_{>}}\, j'_\ell(z_{<})
\,h^{(1)}_\ell(z_{>})
+{z_{>} \over z_{<}}\, j_\ell(z_{<})
\,{h^{(1)}}'_\ell(z_{>})\Biggl)
+O\left(\epsilon^2,|z-z'|^3\right)\Biggl]P_{\ell}(\mu)
\,,
\end{eqnarray}
where $z=\omega r$, $z'=\omega r'$ and $\epsilon=2M\omega$, as before.
Thus the 1PM direct part of the radial Green function
is found as
\begin{eqnarray}
G^{\rm dir}_{\ell m\omega}(r,r')
&&=i\omega\,\Biggl[j_\ell(z_{<})h^{(1)}_\ell(z_{>})
+{\epsilon\over2}\,\Biggl(
\left({1\over z_{<}}+{1\over z_{>}}\right)
j_\ell(z_{<})\,h^{(1)}_\ell(z_{>})\nonumber \\
&& \quad \quad \quad \quad
+{z_{<} \over z_{>}}\, j'_\ell(z_{<})
\,h^{(1)}_\ell(z_{>})
+{z_{>} \over z_{<}}\, j_\ell(z_{<})
\,{h^{(1)}}'_\ell(z_{>})\Biggl)
+O\left(\epsilon^2,|z-z'|^3\right)\Biggr]\,.
\label{1PMdirG}
\end{eqnarray}
The 1PM tail is obtained by subtracting the above from
the full 1PM Green function, (\ref{1PMG}).

Further performing the post-Newtonian expansion, 
the 1PN direct Green function is obtained explicitly as
\begin{eqnarray}
G^{\rm dir}_{\ell m\omega}(r,r') &=& {1 \over 2\ell+1}\,
{1 \over r'}
\left({r \over r'}\right)^{\ell}
\nonumber \\
&&\times\left(
1 +  {\displaystyle \frac {M \,( \ell + 1 )}{r'}}  +
{\displaystyle \frac {1}{2}}
\,{\displaystyle \frac {\omega^{2} r'^{2}}{2\,\ell - 1}}
 - {\displaystyle \frac {M \,\ell}{r}}  - {\displaystyle
\frac {1}{2}} \,{\displaystyle \frac {\omega^{2} r^{2}}{2\,\ell + 3}}
+O(|r-r'|^3)\right)
\,.
\label{1PNdirG}
\end{eqnarray}
Expanding the full 1PM Green function (\ref{1PMG}) to 1PN order,
and subtracting the above direct part (\ref{1PNdirG}) from the result,
we find
\begin{eqnarray}
{}_sG^{\rm tail}_{\ell m\omega}(r,r')=
{s^2 \over 2\ell+1}{1 \over r'} \left({r \over r'}\right)^{\ell}
\left(-{\displaystyle \frac {M}{(\ell + 1)\,r'}}  +
{\displaystyle \frac {M}{\ell\,r}}
\right)\,.
  \label{1PNtailG}
\end{eqnarray}\
It is worthwhile to note that the 1PN tail comes solely from the
$s$-dependent part of the Regge-Wheeler equation.
Thus the scalar d'Alenbertian $\nabla^\mu\nabla_\mu$
contains no 1PN tail, but only the correction to the light cone.
In particular, this implies the 1PN tail is absent for a particle with
scalar charge, in accordance with the recent result of Burko, Liu and
Soen\cite{BuLiSo}.


\section{Regularization}\label{sec:RV}

In this section, we give the procedure to calculate the
regularized Weyl scalar, the tail part of the metric perturbation and
the gravitational reaction force at the position of a point particle.

\subsection{Tail part of the metric perturbation}

{}From the homogeneous solutions $X_\ell^{\rm in}$ and $X_\ell^{\rm up}$
of the Regge-Wheeler equation,
we obtain the corresponding solutions of the Teukolsky equation
by the Chandrasekhar transformation (\ref{eq:CX}).
Using the method described in the previous section,
the radial part of the regularized Green function for the Weyl scalar
${}_{s}{\cal G}_{lm\omega}(r,r')$ is found as
\begin{eqnarray}
{}_{s}{\cal G}_{\ell m\omega}(r,r') =- 2\,\pi\,
{W(X_\ell^{\rm up},X_\ell^{\rm in})
\over W(R_\ell^{\rm up},R_\ell^{\rm in})}
\chi_\ell^{\rm in}\chi_\ell^{\rm up}
\,C_{\omega}(r)\,C_{\omega}(r')\,
{}_s G^{\rm tail}_{\ell m\omega}(r,r')
\,,
\label{Gteuk}
\end{eqnarray}
where $s=-2$ and
$C_{\omega}(r)$ is the Chandrasekhar operator defined by
Eq.~(\ref{Chop}). The factor $-2\pi$ is inserted in the above formula
for ${}_s{\cal G}_{\ell m\omega}$, so that
the Weyl scalar $\psi_4$ is given in the form 
\begin{eqnarray}
  \label{psi4gdef}
  r^4\psi_4(x)=
\int{d\omega\over2\pi}\sum_{\ell m}
\left[{\int dr'{}_{-2}{\cal G}_{\ell m\omega}(r,r')\,
T_{\omega\ell m}(r')r'{}^{-2}(r'-2M)^{-2}}\right]
{}_{-2}Y_{\ell m}(\Omega)e^{-i\omega t}\,,
\end{eqnarray}
with the source term $T_{\omega \ell m}$ defined
in Appendix \ref{append:note}, in accordance with the definition
of Poisson and Sasaki\cite{PoiSas}.

Once we have the regularized Green function for the Weyl scalar,
the regularized metric perturbation can be constructed by
the Chrzanowski transformation\cite{Chrz}.
In the present case, we calculate the tail part (i.e., the regularized
part) of the metric perturbation as follows. We decompose the radial
part of the regularized Green function for the metric perturbation into
parts having different spin weights as
\begin{eqnarray}
G_{(s,s')}^{\ell m\omega}(r,r')
= {32{}_sp_\ell\,{}_{s'}p_\ell\over({}_0p_\ell)^2}
{}_{s}d_{\omega}(r)\,
{}_{s'}D_{\omega}^{\dagger}(r')\,
{}_{-2}{\cal G}_{\ell m\omega}(r,r') \,,
\end{eqnarray}
where the coefficients ${}_sp_\ell$, and the operators
${}_{s}d_{\omega}$ and ${}_{s'}D_{\omega}^{\dagger}$
are defined in Eqs.~(\ref{pelldef}), (\ref{dopdef})
and (\ref{Dopdef}), respectively, of Appendix \ref{append:Chrz}.
Then the tail part of the metric perturbation is expressed as
\begin{eqnarray}
h_{\mu\nu}(\ell m\omega;x) &=&
\Bigl\{l_\mu l_\nu \left( \sum_{s} \int_{2M}^{\infty}
dr'r'{}^{-2}(r'-2M)^{-2} G_{(0,s)}^{\ell m\omega}(r,r')
{}_{s} \overline{T}_{\omega \ell m}(r') \right)
{}_{0}Y_{\ell m}(\theta,\phi)
e^{-i\omega t}
\nonumber \\ &&
-(l_\mu \bar m_\nu +\bar m_\mu l_\nu)
\left( \sum_{s} \int_{2M}^{\infty}
dr'r'{}^{-2}(r'-2M)^{-2} G_{(1,s)}^{\ell m\omega}(r,r')
{}_{s} \overline{T}_{\omega \ell m}(r') \right)
{}_{1}Y_{\ell m}(\theta,\phi)
e^{-i\omega t}
\nonumber \\ &&
+\bar m_\mu \bar m_\nu
\left( \sum_{s} \int_{2M}^{\infty}
dr'r'{}^{-2}(r'-2M)^{-2} G_{(2,s)}^{\ell m\omega}(r,r')
{}_{s} \overline{T}_{\omega \ell m}(r') \right)
{}_{2}Y_{\ell m}(\theta,\phi)e^{-i\omega t}
\Bigr\} \,,
\end{eqnarray}
where $\{l_{\alpha}\}=\{-1, r/(r-2M),0,0\}$ and
$\{\bar m_{\alpha}\}=\{0,0,1,-i \sin \theta \}r/\sqrt{2}$
are the Kinnersley null tetrad,
and ${}_{s} \overline{T}_{\omega \ell m}$
is the complex conjugate of ${}_{s} T_{\omega \ell m}$.
We note that the above metric perturbation is given under
the ingoing radiation gauge condition, defined by
\begin{eqnarray}
h_{\mu\nu}\,l^\mu = 0 \,, \quad
h_{\mu}{}^{\mu} = 0 \,.
\end{eqnarray}

It should be noted that the Chandrasekhar transformation and the
subsequent Chrzanowski transformation in their original forms can be
applied only in the formal sense in the above procedure.
In practice, one should modify those transformations as discussed in
Appendix \ref{append:FGF}, in order to
subtract out the direct part induced by application of the original
transformations.
Let us explain the reason.
The Chandrasekhar transformation involves the second order
derivatives. If one takes the second-order derivatives of the Green
function for the Regge-Wheeler equation in the spacetime domain,
the step function $\theta(-\sigma)$ in front of the tail part is
transmuted to the direct part of the Green function for the Weyl scalar.
This phenomenon shows up in the Fourier-harmonic domain, as demonstrated
in Appendix \ref{append:FGF}. Thus one should modify the Chandrasekhar
transformation so as to subtract this induced direct part. 
As the Chandrasekhar
transformation, the Chrzanowski transformation also involves 
the second-order radial derivatives. Hence, it is necessary to
introduce a modified Chrzanowski transformation to subtract the
induced direct part. In the actual calculation of the metric
perturbation (or the reaction force), however, we do not introduce
the modified Chandrasekhar and Chrzanowski transformations individually,
but construct operators that contain only the first-order derivatives
and that directly transform the Green function for the Regge-Wheeler
variable to that for the components of the metric perturbation (or the
reaction force) as illustrated at the end of Appendix \ref{append:FGF}.
The explicit expressions for the transformation operators are given in
Appendix \ref{append:Chrz}. We note that
thus constructed operators are same as the original operators with
higher derivatives when applied to the full Green function.

When we take the limit in which the position coincides with the particle,
the following facts should be kept in mind.
Let us denote the position of a particle by $\{z^{\alpha}\}
=\{t_0,r_0,\theta_0,\phi_0\}$ and a field point by
$\{x^\mu\}=\{t,r,\theta,\phi\}$.
Because of the causal nature of the tail part, strictly speaking, the
coincidence limit has to be taken by first taking the spatial limit as
$r \rightarrow r_0$, and then taking the limit $t \rightarrow t_0$.
On the other hand, the coincidence limit for the angular coordinates
may be taken at any time, as the metric perturbation depends on the
angular coordinates only through the regular functions
 ${}_{s}Y_{\ell m}$.
But as long as we employ the post-Newtonian (PN) expansion,
we may take any order of the coincidence limit, because the light cone
flattens out in the PN expansion.
Therefore, when we evaluate a quantity in the coincidence limit,
we take the limit $t\to t_0$ first, followed by the spatial coincidence
limit within the context of the PN expansion. After taking the limit,
we perform the summation over $\ell$, $m$ and $\omega$.

Here it is worthwhile to note the following. If we take the mode
summation over $\ell$, $m$ and $\omega$ before we take the spatial
coincidence limit, the result depends on whether the radial limit is
taken from $r$ smaller than $r_0$ or larger than $r_0$. This is a
reflection of the fact that the first-order radial derivatives remaining
in the final operators that transform the Regge-Wheeler variable to the
metric perturbation (or to the reaction force) induce a `quasi-direct'
part that has a direction-dependent limit. Nevertheless, the
avarage of the two different limits turns out to be the same as the
limit obtained by first taking the coincidence limit before the mode
summation. This can be explained by considering the first derivative of
the step function in the spacetime domain. It has a 
direction-dependent limit, but this difference 
disappears when averaged over all directions.


\subsection{Force for a circular orbit}

The form of the reaction force which describes the deviation from the
geodesic motion on the background is given by Eq.~(\ref{eq:force}).
Here we focus on circular motion.
Since only the $t$- and $\phi$-components of the four velocity are
non-zero in this case, the radial component of the reaction force can be 
expressed as
\begin{eqnarray}
F^{r}(x) = -{1 \over 2} g^{rr}&& \Bigl[
\left(2h_{rt;t}-h_{tt;r}\right)u^{t}u^{t}
\nonumber \\ &&
+\left(2h_{r\phi;\phi}-h_{\phi\phi;r}\right)
u^{\phi}u^{\phi}
+2 \left(h_{tr;\phi}+h_{\phi r;t}
- h_{t\phi;r}\right)
u^t u^{\phi}\Bigr]_{\rm tail} \,,
\end{eqnarray}
where $[\cdots]_{\rm tail}$ indicates that
$h_{\mu\nu;\alpha}$ should be interpreted as
the tail part $h^{\rm tail}_{\mu\nu;\alpha}$.
Specializing to the case of Schwarzschild background,
we have
\begin{eqnarray}
F^{r}(x) = -{1 \over 2} g^{rr}&& \Bigl[
\left(2h_{rt,t}-h_{tt,r}-{2\,M\,f \over r^2}h_{rr} \right)
u^{t}u^{t}
\nonumber \\ &&
+\left(2h_{r\phi,\phi}-h_{\phi\phi,r}
+2\,r \sin ^2 \theta \,f h_{rr}\right)
u^{\phi}u^{\phi}
\nonumber \\ &&
+2 \left(h_{tr,\phi}+h_{\phi r,t}
- h_{t\phi,r}\right)
u^t u^{\phi} \Bigr]_{\rm tail} \,.
\label{Fsch}
\end{eqnarray}
Since there is no effect of the gravitational radiation at 
1PN order, we may consider only the radial component of the reaction
force to that order.
 Physically, the radial component describes the correction to the
radius of the orbit that deviates from the geodesic in the unperturbed
background.

As discussed in the previous subsection, we construct
the operators that directly trasform the Regge-Wheeler variable into
the reaction force expressed in Eq.~(\ref{Fsch}). At 1PN order,
we find that only the $h_{tt,r}^{\rm tail}$ term contributes.
Since this term involves the operation $\partial_r{}_0\Delta_{\omega}$,
where ${}_0\Delta_{\omega}$ is defined in Appendix \ref{append:Chrz},
Eq.~(\ref{eq:del0}), we reduce the second order derivative
in it by using the homogeneous Regge-Wheeler equation, as discussed in
the preveous subsection.
The result is
\begin{eqnarray}
\partial_r \, {}_0\Delta_{\omega} &=&
\left(-{1 \over 4} \ell (\ell+1)\left(1-{2M \over r}\right)
- i \, \omega r\left(1-{9M \over 2r}\right)
+{1 \over 2} \omega^2 r^2 + {M \over r}\left(1-{3M \over r}\right)
\right){1 \over r}\partial_r
\nonumber \\ &&
+{1 \over 4 r^2}\biggl(- \ell (\ell+1)\left(1 + 2\,i\,\omega r
-{6M \over r}\right) + 2 \omega^2 r^2 \left(2+ i\,\omega r
-{9M \over r} \right)
\nonumber \\ && \quad \qquad
-{4 M^2 \over r^2}\left(2\,\ell^2+2\,\ell+3 -{6M \over r}\right)
+ 2\,i\,\omega \left(2\ell^2+2\ell+5-{12M \over r}\right)
\biggr) \,.
\end{eqnarray}
This operator is used to derive the 1PN reaction force term.

After summing over $\ell, m$ and $\omega$,
and taking the coincidence limit,
the gravitational reaction force to 1PN order
is found to be
\begin{eqnarray}
F^{r}(z) = {11\,M\mu \over 3\,r_0^3} \,.
\label{Freact}
\end{eqnarray}
We warn that this result by itself has no physical significance.
First, it contains contributions only from harmonics with
$\ell\geq2$. We have to calculate the $\ell=0$ and $1$
components to make the force complete. Second, it is obtained in the
ingoing radiation gauge, and since we lack a method to calculate the
gravitational radiation to second order in $\mu$, the only thing we
can do at present is to compare the result with the standard 1PN formula
given in the literature.  To do so, we have to make a gauge
transformation from the ingoing radiation gauge to the harmonic gauge.
Only after this procedure can we determine if our result is valid.


\section{Discussion}\label{sec:discussion}

In this paper we have proposed a new method
to derive the gravitational reaction force on a point particle
in the Schwarzschild background.
In this method, the tail part of the metric perturbation, which is
responsible for the reaction force, is obtained by first
considering the Fourier-harmonic expansion of the Green function for the
Regge-Wheeler equation. After calculating the tail part of the Green
function, the result is transformed
to the metric perturbation through the Chandrasekhar transformation
followed by the Chrzanowski transformation. Since the extraction of the
tail part is done at the level of the Green function in
the Fourier-harmonic expanded form, we call our method the mode-by-mode
regularization of the Green function.
An advantage of our method is that it is relatively easy to obtain the
tail part of the Green function for the Regge-Wheeler equation if we
adopt the post-Minkowskian (or post-Newtonian) expansion.

However, there are still a couple of issues to be resolved. One is that
the $\ell=0$ and 1 components are not included in the metric
perturbation obtained with the Chrzanowski transformation. Hence the tail
part in them must be derived by some other method. The $\ell=0$ and 1
components of the full metric perturbation can be solved exactly in the
Regge-Wheeler-Zerilli gauge, as shown by Zerilli\cite{Zer}.
Therefore, what we need is a method to identify the tail part in the
$\ell=0$ and 1 components.
The other, most crucial issue is that the physical meaning of the
resulting reaction force remains unclear, since the reaction force is a
gauge-dependent notion.
For an ultimate resolution of this essential problem, we need to
develop a second-order perturbation theory so as to make it possible
to calculate the $O(\mu)$ correction to the gravitational radiation.

Nevertheless, at low post-Newtonian orders, there is a way to resolve 
these issues. The equations of motion for a two-body system and the
gravitational radiation from such a system are investigated extensively
in \cite{BIWW,WilWis,BFP,BlaFay} 
in the harmonic gauge. Hence, by performing the gauge
transformation to the harmonic gauge, we should be able to
clarify the physical effect of our result, as well as to examine
the consistency of our result with those of these previous works. 
As a first step, we are
currently working on a consistency check at the 1PN order\cite{NakSas}.


\noindent

\section*{Acknowledgements}

We thank U. Gen, Y. Himemoto and N. Sago for fruitful conversations.
Special thanks are due to Y. Mino, H. Tagoshi and T. Tanaka
for invaluable discussions.
This work was supported in part by a Monbusho Grant-in-Aid
for Creative Research (No.~09NP0801), and by
a Monbusho Grant-in-Aid for Scientific Research (No.~12640269).
H.N. is supported by a Research Fellowships of the
Japan Society for the Promotion of Science
for Young Scientists (No.~2397).


\appendix

\section{The Homogeneous 1PM Regge-Wheeler Functions}\label{1PMX}
Here we recapitulate the formulas for the homogeneous
1PM Regge-Wheeler radial functions $X_\ell^{\rm in}$ and
$X_\ell^{\rm up}$, which can be found in Poisson and Sasaki\cite{PoiSas}
and Leonard and Poisson\cite{LeoPoi}.

The formulas are
\begin{eqnarray}
X_\ell^{\rm in}(z)
&&=2{\cal A}_{\ell\omega}^{\rm in}
(-i)^{\ell+1}e^{i\epsilon(\ln(2\epsilon)-\beta_\ell)}
(1+{\pi\over2}\,\epsilon)\,z
\left[j_\ell(z)+\epsilon\,\xi_\ell^{\rm in}(z)+O(\epsilon^2)\right],
\nonumber\\
&&\xi_\ell^{\rm in}(z)=
-A_\ell(z)j_\ell(z)+B_\ell(z)n_\ell(z)
-{\ell^2-s^2\over2\ell(2\ell+1)}j_{\ell-1}(z)
+{(\ell+1)^2-s^2\over2(\ell+1)(2\ell+1)}j_{\ell+1}(z),
\label{1PMXin}\\
X_\ell^{\rm up}(z)
&&=i^{\ell+1}e^{-i\epsilon(\ln(2\epsilon)-\beta_\ell)}
(1-{\pi\over2}\,\epsilon)\,z
\left[h^{(1)}_\ell(z)+\epsilon\,\xi_\ell^{\rm up}+O(\epsilon^2)\right],
\nonumber\\
&&\xi_\ell^{\rm up}(z)=
(A_\ell(z)-iB_\ell(z))h^{(1)}_\ell(z)+C_\ell(z)j_\ell(z)
-{\ell^2-s^2\over2\ell(2\ell+1)}h^{(1)}_{\ell-1}(z)
+{(\ell+1)^2-s^2\over2(\ell+1)(2\ell+1)}h^{(1)}_{\ell+1}(z),
\label{1PMXup}
\end{eqnarray}
where $\epsilon=2M\omega$, $z=\omega r$, and the functions
$A_\ell(z)$, $B_\ell(z)$ and $C_\ell(z)$
are given by
\begin{eqnarray}
  A_\ell(z)&=&\Si(2z)+z^2n_0(z)j_0(z)
+\sum_{p=1}^{\ell-1}\left({1\over p}+{1\over p+1}\right)z^2n_p(z)j_p(z)\,,
\nonumber\\
 B_\ell(z)&=&\Ci(2z)-\gamma-\ln(2z)+z^2j_0(z)^2
+\sum_{p=1}^{\ell-1}\left({1\over p}+{1\over p+1}\right)z^2j_p(z)^2\,,
\nonumber\\
 C_\ell(z)&=&2\left({\pi\over2}-A_\ell(z)
 +i\left(B_\ell(z)+\gamma+\ln(2z)\right)\right)
-i\left(1+\sum_{p=1}^{\ell-1}\left({1\over p}+{1\over p+1}\right)z
(-1)^p R_{2p,{1\over2}-p}(z)\right),
  \label{ABCdef}
\end{eqnarray}
for $\ell\geq1$, and
\begin{eqnarray}
 && A_0(z)=\Si(2z)\,,
\nonumber\\
 && B_0(z)=\Ci(2z)-\gamma-\ln(2z)\,,
\nonumber\\
 && C_0(z)=2\left({\pi\over2}-\Si(2z)+i\Ci(2z)\right),
\end{eqnarray}
for $\ell=0$. It should be noted that the expressions~(\ref{1PMXin}) and
(\ref{1PMXup}) are valid only for $\omega\ge0$. For $\omega<0$
(i.e, for $z<0$), they should be replaced by
$\overline{X^{\rm in}_\ell(|z|)}$ and $\overline{X^{\rm up}_\ell(|z|)}$,
respectively.

\section{The Chrzanowski Transformation}\label{append:Chrz}

In this appendix, we summarize the Chrzanowski method\cite{Chrz}
to construct the metric pertubation from the Weyl scalar and describe
the modification of differential operators necessary to extract the
tail part of the metric perturbation.

Using the method of Chrzanowski\cite{Chrz},
we can generate metric perturbations
in either the in-going radiation gauge or the out-going radiation
gauge from the Weyl scalar.
We first introduce the following differential operators
on the Kerr spacetime:
\begin{eqnarray}
\hat h^{\rm in}_{nn}&=&
-(\bar \delta+\alpha+3\bar \beta-\bar \tau)
(\bar \delta+4\bar \beta+3\bar \tau) \,,
\nonumber\\
\hat h^{\rm in}_{mm}&=&
-(D-\bar \rho+3\bar \epsilon-\epsilon)(D+3\bar \rho+4\bar \epsilon) \,,
\nonumber\\
\hat h^{\rm in}_{nm}&=&-{1/2}
\bigl\{(D+\rho-\bar \rho+\epsilon+3\bar \epsilon)
(\bar \delta+4\bar \beta+3\bar \tau)
\nonumber \\ && \qquad \qquad
+(\bar \delta+3\bar \beta-\alpha-\pi-\bar \tau)
(D+3\bar \rho+4\bar \epsilon)\bigr\} \,,
\label{hinop}
\end{eqnarray}
and
\begin{eqnarray}
\hat h^{\rm out}_{ll}&=&
-(\delta+\bar \pi-3\bar \alpha-\beta)(\delta-3\bar \pi-4\bar \alpha)
(\bar \rho)^{-4} \,,
\nonumber\\
\hat h^{\rm out}_{\bar m \bar m}&=&
-(\Delta+\gamma-3\bar \gamma+\bar \mu)(\Delta-4\bar \gamma-3\bar \mu)
(\bar \rho)^{-4} \,,
\nonumber \\
\hat h^{\rm out}_{l\bar m}&=&-{1/2}
\bigl\{(\delta+\tau+\bar \pi-3\bar \alpha+\beta)
(\Delta-4\bar \gamma-3\bar \mu)
\nonumber \\ && \qquad \qquad
+(\Delta-\gamma-3\bar \gamma+\bar \mu-\mu)
(\delta-3\bar \pi-4\bar \alpha)\bigr\}
(\bar \rho)^{-4} \,.
\label{houtop}
\end{eqnarray}
Here $D=l^\mu\nabla_\mu$,
$\Delta=n^\mu\nabla_\mu$, $\delta=m^\mu\nabla_\mu$
and $\bar\delta=\bar m^\mu\nabla_\mu$, 
where $\{\ell^\mu$, $n^\mu$, $m^\mu$, $\bar m^\mu\}$ is 
the Kinnersley null tetrad,
$\rho$, $\mu$, $\tau$, $\pi$, $\epsilon$, $\gamma$,
$\alpha$ and $\beta$ are the spin coefficients
of the Newman-Penrose formalism\cite{NewPen},
and the overline indicates complex conjugation.

With these differential operators,
we have the following metric perturbation in the Fourier-harmonic expansion
in the ingoing and outgoing radiation gauges:
\begin{eqnarray}
h^{\rm in}_{\mu\nu}(\ell m\omega)&=&
\bigl\{l_\mu l_\nu \hat h^{\rm in}_{nn}
+\bar m_\mu \bar m_\nu \hat h^{\rm in}_{mm}
-(l_\mu \bar m_\nu +\bar m_\mu l_\nu)\hat h^{\rm in}_{nm}\bigr\}
{}_{-2}R_{\ell m\omega}(r){}_{2}Y_{\ell m}(\theta,\phi)e^{-i\omega t}\,,
\label{eq:ingoingxx}
\\
h^{\rm out}_{\mu\nu}(\ell m\omega)&=&
\bigl\{n_\mu n_\nu \hat h^{\rm out}_{ll}
+m_\mu m_\nu \hat h^{\rm out}_{\bar m \bar m}
-(n_\mu m_\nu +m_\mu n_\nu)\hat h^{\rm out}_{l \bar m}\bigr\}
{}_{2}R_{\ell m\omega}(r){}_{-2}Y_{\ell m}(\theta,\phi)e^{-i\omega t}\,.
\label{eq:outgoingxx}
\end{eqnarray}
Here
${}_sR_{lm\omega}(r)$ is the Teukolsky function of spin index $s$, and
${}_sY_{\ell m}(\theta,\phi)$ is the spin weighted spherical
harmonic function.

Now we specify
the background spacetime to be a Schwarzschild blackhole.
We assume operands are expanded in the Fourier-harmonic functions.
We use the pure radial and angular operators
introduced by Chandrasekhar\cite{Chantrans},
\begin{eqnarray}
{\cal D}_n &=&
\partial_r -{ir^2\omega \over\Delta_{K}}+2n{r-M\over \Delta_{K}}
\,, \quad
{\cal D}_n^\dagger =
\partial_r +{ir^2\omega \over\Delta_{K}}+2n{r-M\over \Delta_{K}}
\,, \\
{\cal L}_n &=& \partial_\theta +m{1\over \sin\theta}+n\cot\theta \,,
 \quad
{\cal L}_n^\dagger=\partial_\theta -m{1\over \sin\theta}+n\cot\theta\,,
\end{eqnarray}
where $\Delta_{K}=r(r-2M)$.
The Newman-Penrose operators and spin coefficients become
\begin{eqnarray}
&& D \,=\, {\cal D}_0 \,, \quad
\Delta \,=\, -{\Delta_{K}\over 2r^2}{\cal D}_0^\dagger \,, \quad
\delta \,=\, {1\over \sqrt 2\,r}{\cal L}_0^\dagger \,, \\
&& \rho \,=\, -{1\over r} \,, \quad
-\alpha \,=\, \beta \,=\, {\cot\theta\over 2\sqrt2\,r} \,, \quad
\mu \,=\, -{\Delta_{K}\over 2r^3} \,, \quad
\gamma \,=\, {M\over 2r^2} \,.
\end{eqnarray}
The radial differential operator ${\cal L}$ introduced
 in Eq.~(\ref{eq:L})
is related to the operator $D$ in the above as ${\cal L}=f D$,
where $f=(r-2M)/r$.
Then all the operators defined in Eqs.~(\ref{hinop}) and (\ref{houtop})
are expressed as
\begin{eqnarray}
&& \hat h^{\rm in}_{nn} \,=\, -{1\over 2 r^2}{\cal L}_1{\cal L}_2 \,, \ \quad
\hat h^{\rm in}_{nm} \,=\, -{1\over \sqrt 2 \,r}\Bigl({\cal D}_0-{2\over r}\Bigr)
{\cal L}_2 \,, \quad
\hat h^{\rm in}_{mm} \,=\, -\Bigl({\cal D}_0 +{1\over r}\Bigr)
\Bigl({\cal D}_0-{3\over r}\Bigr) \,, \\
&& \hat h^{\rm out}_{ll} \,=\, -{r^2 \over 2}
{\cal L}^\dagger_1{\cal L}^\dagger_2 \,, \quad
\hat h^{\rm out}_{l\bar m} \,=\, {r\Delta_{K}\over 2\sqrt 2}
\Bigl({\cal D}^\dagger_2-{2\over r}\Bigr){\cal L}^\dagger_2 \,, \quad
\hat h^{\rm out}_{\bar m\bar m} \,=\, -{\Delta_{K}^2\over 4}
\Bigl({\cal D}^\dagger_2+{1\over r}\Bigr)
\Bigl({\cal D}^\dagger_2-{3\over r}\Bigr) \,.
\end{eqnarray}

In the remainder of this appendix, operators used to derive
the metric perturbation from the Regge-Wheeler variable are
given. We consider the metric perturbation in the ingoing radiation
gauge.

First, note that, when operating on a quantity of spin weight $s$, the
angular operators ${\cal L}_s$ and ${\cal L}_{-s}^\dagger$ are just the
``edth'' operators given in Eq.~(\ref{edthdef}), except for the over-all
sign:
\begin{eqnarray}
  \edth=-{\cal L}_{-s}^\dagger\,,\quad
  \bar\edth=-{\cal L}_s\,.
\end{eqnarray}
Hence
\begin{eqnarray*}
{\cal L}_2 \,({}_2 Y_{\ell m}) &=&
-\bar\edth\,({}_2 Y_{\ell m})
= [(\ell -1)(\ell +2)]^{1/2}{}_1 Y_{\ell m}
\,, \\
{\cal L}_1{\cal L}_2 \,({}_2 Y_{\ell m}) &=&
\bar\edth^2\,({}_2 Y_{\ell m})
= [(\ell -1) \ell (\ell +1)(\ell +2)]^{1/2}{}_0 Y_{\ell m}
\,.
\end{eqnarray*}
Accordingly, we put
\begin{eqnarray}
\hat h^{\rm in}_{nn} &=& {}_0 p_{\ell} \,
{}_{0}\hat d_{\omega} \,,
\nonumber\\
\hat h^{\rm in}_{nm} &=& {}_{-1} p_{\ell} \,
{}_{1}\hat d_{\omega} \,,
\nonumber \\
\hat h^{\rm in}_{mm} &=& {}_{-2} p_{\ell} \,
{}_{2}\hat d_{\omega} \,,
\end{eqnarray}
where the coefficients ${}_s p_{\ell}$ ($s=0,-1,-2$) are defined by
\begin{eqnarray}
{}_{0}p_{\ell} &=&
      2[(\ell -1) \ell (\ell +1)(\ell +2)]^{1/2}\,,
\nonumber\\
{}_{-1}p_{\ell} &=&
      2[2(\ell -1)(\ell +2)]^{1/2}\,,
\nonumber\\
{}_{-2}p_{\ell} &=&
      1\,,
\label{pelldef}
\end{eqnarray}
and ${}_{s}\hat d_{\omega}$ $(s=0,1,2)$
are the radial differential operators defined by
\begin{eqnarray}
{}_{0}\hat d_{\omega} &=& -{1\over 4\,r^2} \,,
\nonumber\\
{}_{1}\hat d_{\omega} &=& -{1\over 4} r\,f^{-1}{\bar{\cal L}}\,
r^{-2} \,,
\nonumber\\
{}_{2}\hat d_{\omega} &=& -r^{-1}f^{-1}{\bar{\cal L}}\,
r^4f^{-1}{\bar{\cal L}}\,r^{-3} \,.
\label{dopdef}
\end{eqnarray}

Next, we rewrite the Green function for the Weyl scalar $\psi_4$
in a more convenient manner, so that the source term is divided
into the tetrad components of the energy-momentum tensor that
have specific spin weights, ${}_sT_{\omega\ell m}$,
as defined by Eq.~(\ref{sTdef}).
For this purpose, we introduce the operators
${}_{s}D_{\omega}$ and ${}_{s}D_{\omega}^{\dagger}$ defined in
 Poisson and Sasaki \cite{PoiSas}:
\begin{eqnarray}
{}_{0} D_{\omega} &=& {}_{0} D_{\omega}^{\dagger}=
r^4 \,,
\nonumber\\
{}_{-1} D_{\omega} &=&
r^2 f {\cal L}\, r^{3} f^{-1} \,,
\nonumber\\
{}_{-2} D_{\omega} &=&
r f\, {\cal L}\, r^4 f^{-1} {\cal L}\, r \,,
\nonumber\\
{}_{-1} D_{\omega}^{\dagger} &=&
-r^7 {\bar{\cal L}}\, r^{-2} \,,
\nonumber\\
{}_{-2} D_{\omega}^{\dagger} &=&
r^5 f\, {\bar{\cal L}}\, r^4 f^{-1} {\bar{\cal L}}\, r^{-3} \,.
\label{Dopdef}
\end{eqnarray}
Using these operators, we can rewrite the expression for $\psi_4$
given by Eq.~(\ref{psi4gdef}) as
\begin{eqnarray}
 r^4\psi_4(x)&=&
\int{d\omega\over2\pi}\sum_{\ell m}
\left[{\int_{2M}^\infty dr'{}_{-2}{\cal G}_{\ell m\omega}(r,r')\,
T_{\omega\ell m}(r')\,r'{}^{-2}(r'-2M)^{-2}}\right]
{}_{-2}Y_{\ell m}(\Omega)e^{-i\omega t}
\nonumber\\
& =&\int{d\omega\over2\pi}\sum_{\ell m}
\left[\sum_s \int_{2M}^{\infty}
 dr'r'{}^{-2}(r'-2M)^{-2}
{}_{(s)}G_{\ell m \omega}(r,r')\,
{}_{s}T_{\omega \ell m}(r') \right]
\, {}_{-2}Y_{\ell m}(\Omega) e^{-i\omega t}\,,
\end{eqnarray}
where the Green Functions ${}_{(s)}G_{\ell m \omega}(r,r')$
 ($s=0,-1,-2$) are defined by
\begin{eqnarray}
{}_{(s)}G_{\ell m \omega}(r,r')
= {}_s p_{\ell} \, {}_{s} D_{\omega}^{\dagger}(r')
\, {}_{-2}{\cal G}_{\ell m\omega}(r,r')\,,
\end{eqnarray}
with ${}_{s} D_{\omega}^{\dagger}(r')$ acting on the $r'$-dependent
part of ${}_{-2}{\cal G}_{\ell m\omega}(r,r')$ .
The relation between the Teukolsky radial function
${}_{-2}R_{\ell m \omega}(r)$ to be used in the Chrzanowski
transformation (\ref{eq:ingoingxx}) and
the radial part of $r^4 \psi_4(x)$ is
\begin{eqnarray}
{}_{-2} R_{\ell m \omega}(r) = {32 \over ({}_0p_{\ell})^2} \,
\sum_s \int_{2M}^{\infty} dr' r'^{-2} (r'-2M)^{-2}
{}_{(s)}G_{\ell m \omega}(r,r') {}_{s} T_{\omega \ell m}(r') \,.
\end{eqnarray}

Here, however, we wish to construct a Green function
that directly gives the metric perturbation.
For this purpose, we introduce the differential operators
${}_s\Delta_\omega={}_sd_\omega\,C_\omega$
and ${}_s\Lambda_\omega={}_sD_\omega^\dagger\,C_\omega$,
 where ${}_sd_\omega$ and ${}_sD_\omega^\dagger$ are
 defined by Eqs.~(\ref{dopdef}) and (\ref{Dopdef}), respectively,
and $C_\omega$ is the Chandrasekhar operator defined by
Eq.~(\ref{Chop}).
Explicitly, ${}_s\Delta_\omega$ and ${}_s\Lambda_\omega$
are given by
\begin{eqnarray}
{}_0 \Delta_{\omega} &=& -{1 \over 4} f\,{\cal L}\,f^{-1}
{\cal L}\,r
= -{1 \over 4} r^{-1}\, {}_0 \Gamma_{\omega} \,,
\label{eq:del0} \\
{}_1 \Delta_{\omega} &=& -{1 \over 4}\,r\,f^{-1}
{\bar{\cal L}}\,f\,{\cal L}\,f^{-1}{\cal L}\,r
= {1 \over 4} r^{-1}f^{-1}\, {}_{-1} \Gamma_{\omega} \,, \\
{}_2 \Delta_{\omega} &=& -r^{-1}f^{-1}{\bar{\cal L}}\,
r^4 f^{-1}{\bar{\cal L}}\,r^{-1}f\,{\cal L} f^{-1}{\cal L}\,r
= - r^{-1} f^{-2}\,{}_{-2} \Gamma_{\omega} \,,
\end{eqnarray}
where the operators ${}_{s} \Gamma_{\omega}$
are introduced in \cite{PoiSas} and are given by
\begin{eqnarray}
{}_0 \Gamma_{\omega} &=& (\omega \, r)^{-1} C_{\omega}
\nonumber \\
&=& 2\left(1-{3\,M \over r}+i\,\omega r \right)r\,f {d \over dr}
+ f\left(\ell (\ell +1)-{6\,M \over r} \right)
+ 2\,i\,\omega r \left(1-{3\,M \over r}+i\,\omega r \right) \,, \\
{}_{-1} \Gamma_{\omega} &=& -r^2 {\bar{\cal L}}\,f\,{\cal L}\,
f^{-1}{\cal L}\,r
\nonumber \\
&=& -f \left[ \left(\ell (\ell +1)+2\,i\,\omega r \right)
r\,f {d \over dr} + \ell (\ell +1)(f+ i\,\omega r)
-2(\omega r)^2 \right] \,, \\
{}_{-2} \Gamma_{\omega} &=& f\,{\bar{\cal L}}\,r^4f^{-1}
{\bar{\cal L}}r^{-1}f\,{\cal L}\,f^{-1}{\cal L}\,r
\nonumber \\
&=& f^2 \left[ 2 \left((\ell -1) (\ell +2)+ {6\,M \over r} \right)
r\,f {d \over dr} + (\ell -1) (\ell +2)(\ell (\ell +1)+ 2\,i\,\omega r)
-12{f\,M \over r} \right] \,.
\end{eqnarray}
With these newly introduced operators, the regularized Green functions
for the components of the metric perturbation can be written as
\begin{eqnarray}
G_{(s,s')}^{\ell m\omega}(r,r') =
-{64\,\pi\over ({}_0p_{\ell})^2}\,
{W(X_\ell^{\rm up},X_\ell^{\rm in})
\over W(R_\ell^{\rm up},R_\ell^{\rm in})}\,
\chi_\ell^{\rm in}\,\chi_\ell^{\rm up}\,
{}_{s}p_{\ell}\,{}_{s'}p_{\ell}\,
{}_{s} \Delta_{\omega}(r)\,
{}_{s'} \Lambda_{\omega}(r')\,
{}_{-2}G_{\ell m \omega}^{\rm tail}(r,r') \,,
\end{eqnarray}
where the coefficeints $\chi_\ell^{\rm in}$ and $\chi_\ell^{\rm up}$
are defined in Eq.~(\ref{chielldef}).
Note that the operators ${}_s \Delta_{\omega}$ and
${}_{s} \Gamma_{\omega}$ defined above become the modified differential
operators by reducing the higher $r$-derivatives using
the homogeneous Regge-Wheeler equation.
As discussed in Sec.~\ref{sec:RV}, these operators transform the tail
of the Regge-Wheeler Green function into the tail of the Green
functions for the metric components. If the original higher differential
operators are used, the direct part is partially induced, as
shown in Appendix \ref{append:FGF}.


\section{The Induced Direct Part of the Green Function}
\label{append:FGF}

In Appendix \ref{append:Chrz},
we introduced modified differential operators
that differ from the original Chandrasekhar and Chrzanowski operators
in order to obtain the tail part of the metric perturbation.
This is because the direct part is partially induced by
taking derivatives of the tail part of the Regge-Wheeler Green function,
and such an induced part must be subtracted out. In this appendix,
we explicitly show this phenomenon at 1PN order.
Then we illustratively summarize our prescription to
subtract the induced direct part from the Green function for the metric
perturbation.

The homogeneous radial Regge-Wheeler function satisfies
\begin{eqnarray}
\left[{\cal O}_0
+ \epsilon\,\left({\cal O}_1+ {s^2 \over z^3}\right)
\right]\,X_{\ell} = 0 \,,
\label{eq:RWH}
\end{eqnarray}
where $z=\omega r$, $\epsilon= 2 \omega M$,
and the operators ${\cal O}_0$ and ${\cal O}_1$ are given by
\begin{eqnarray}
{\cal O}_0 && = {d^2 \over dz^2} + 1 - {\ell (\ell+1) \over z^2} \,,
\nonumber\\
{\cal O}_1 &&= {2 \over z}{d^2 \over dz^2}
+ {1 \over z^2}{d \over dz} + {\ell (\ell+1) \over z^3}
-{1 \over z^3} \,.
\end{eqnarray}
The Green function is constructed from the homogeneous functions as
$X_\ell^{\rm in}(z)X_\ell^{\rm up}(z')$. Therefore we consider
this form.
The homogeneous Regge-Wheeler functions can be expanded to 
1PM order as
\begin{eqnarray}
X_\ell = X_0 + \epsilon \left(X_1^d + s^2 X_1^t\right) + O(\epsilon^2)\,,
\label{eq:RWHF}
\end{eqnarray}
where $X_1^d$ and $X_1^t$ are the parts of the 1PM correction
independent of $s$ and dependent on $s$, respectively.
When we further adopt the post-Newtonian expansion,
they correspond to the parts of the homogeneous
function that contribute to the direct and tail parts
of the Green function, respectively, as discussed in
Sec.~\ref{sec:1PMtail}.
Specifically, the direct part of the Green function consists of
$X_0(z)X_0(z')$  and $X_0(z)X_1^d(z')+X_1^d(z)X_0(z')$, and the tail
part consists of $X_0(z)X_1^t(z')+X_1^t(z)X_0(z')$.

We substitute Eq.~(\ref{eq:RWHF}) into Eq.~(\ref{eq:RWH}) and
collect terms of the same order in $\epsilon$. Then we find 
\begin{eqnarray}
&&{\cal O}_0 \,X_0 = 0 \,,
\nonumber\\
&&{\cal O}_1\,X_0 +{\cal O}_0\,X_1^d
+s^2 \left({\cal O}_0 \, X_1^t+{1 \over z^3}\,X_0\right) = 0 \,.
\label{eq:s-s}
\end{eqnarray}
Since $X_0$ and $X_1^d$ are independent of $s$,
it follows that
\begin{eqnarray}
{\cal O}_1\,X_0+{\cal O}_0 \,X_1^d & =& 0 \,,
\nonumber\\
{\cal O}_0 \, X_1^t+{1 \over z^3}\,X_0 & =& 0 \,.
\label{eq:t-d}
\end{eqnarray}
We find from the second equation of the above that
the tail part of the Regge-Wheeler Green function
is transformed into the direct part by the operation
${\cal O}_0$.

To subtract out such an induced direct part,
it is therefore nessesary to reduce the second-order derivative
by the equation ${\cal O}_0 X_1^t=0$ whenever second- or higher-order
derivatives operate on the Regge-Wheeler Green function.
Although we do not have a general proof, we expect the same
to be true at any order of the post-Newtonian expansion.
Thus, in general, the prescription is to reduce
the higher-order derivatives on the tail part by the
equation
\begin{eqnarray}
\left[{\cal O}_0
+ \epsilon\,\left({\cal O}_1+ s^2\,{1 \over z^3}\right)
\right]\,X_{\ell}^t = 0 \,,
\end{eqnarray}
where $X_\ell^t$ is the part of the homogeneous Regge-Wheeler
funtion that constitutes the tail part of the Green function.

It may be useful to summarize the above prescription illustratively.
First, one constructs the full Regge-Wheeler Green function, say
$G_{RW}$. Then the full Green function for the metric
perturbation, say $G$, is obtained as
\begin{eqnarray}
  G(x,x')
={\cal O}(x){\cal O}'(x')G_{\rm RW}(x,x'),
\end{eqnarray}
where ${\cal O}$ and ${\cal O}'$ are some higher-order differential operators.
As long as $x\neq x'$, one can then reduce the higher-order derivatives
in ${\cal O}$ and ${\cal O}'$  by using the homogeneous Regge-Wheeler
equation without any problem. Let us denote the reduced operators 
thus obtained 
by $\tilde{\cal O}$ and $\tilde{\cal O}'$.
Then we have
\begin{eqnarray}
  G(x,x')
=\tilde{\cal O}(x)\tilde{\cal O}'(x')G_{\rm RW}(x,x').
\end{eqnarray}
 By construction, the operators $\tilde{\cal O}$ and $\tilde{\cal O}'$
contain only the first derivatives.
Now we divide $G_{\rm RW}$ into the direct part $G_{\rm RW}^{\rm dir}$
and the tail part $G_{\rm RW}^{\rm tail}$.
 Then the tail part of the Green function for the metric
is given by
\begin{eqnarray}
  G^{\rm tail}(x,x')
=\tilde{\cal O}(x)\tilde{\cal O}'(x')G_{\rm RW}^{\rm tail}(x,x').
\end{eqnarray}
This is our prescription to obtain the tail part of the
Green function for the metric perturbation.


\section{Notation for the Source Term}\label{append:note}

The source term of the inhomogeneous Teukolsky equation
is constructed from the energy-momentum tensor of a point particle:
\begin{eqnarray}
T^{\alpha\beta}(x) = \mu \int d\tau u^{\alpha} u^{\beta}
\delta^{(4)}[x-x'(\tau)] \,.
\end{eqnarray}
The first step is to obtain the projections by the null tetrad,
$T_{\alpha\beta}\,n^{\alpha}\,n^{\beta}$,
$T_{\alpha\beta}\,n^{\alpha}\,\bar{m}^{\beta}$ and
$T_{\alpha\beta}\,\bar{m}^{\alpha}\,\bar{m}^{\beta}$.
Then we calculate the Fourier-harmonic components according to
\begin{eqnarray}
{}_s T_{\omega \ell m} =
{1 \over 2\pi} \int dt\,d\Omega\,{}_{s}T(x) \,
\overline{{}_{s} Y_{\ell m}(\Omega)} \,e^{i \omega t} \,,
\label{sTdef}
\end{eqnarray}
where the quantities ${}_sT$ ($s=0,-1,-2$) are defined by
\begin{eqnarray}
  {}_0T=T_{\alpha\beta}\,n^{\alpha}\,n^{\beta}\,,
\quad
{}_{-1}T=T_{\alpha\beta}\,n^{\alpha}\,\bar{m}^{\beta}\,,
\quad
{}_{-2}T=T_{\alpha\beta}\,\bar{m}^{\alpha}\,\bar{m}^{\beta}\,.
\end{eqnarray}
Then the source term $T_{\omega \ell m}$ for the radial Teukolsky
equation is obtained by operating on ${}_sT_{\omega \ell m}$ 
with relevant diferential operators 
and then by summing over $s$ as 
\begin{eqnarray}
T_{\omega \ell m}
= \sum_{s} {}_s p_{\ell} \,
{}_s D_{\omega} \,{}_s T_{\omega \ell m} \,,
\end{eqnarray}
where the ${}_sD_{\omega}$ are the differential operators defined in
Eq.~(\ref{Dopdef}).

We note that the source ${}_{-2}{\cal T}$ of the Teukolsky
equation (\ref{teuk}) is related to the above $T_{\omega\ell m}$ as
\begin{eqnarray}
{}_{-2}{\cal T}_{\omega\ell m}
={1 \over 2\pi} \int dt\,d\Omega \,
{}_{-2}{\cal T}(x)\,
\overline{{}_{-2} Y_{\ell m}(\Omega)} \,e^{i \omega t}
={T_{\omega \ell m}\over 2\,r^2}\,.
\end{eqnarray}
In this paper, we use the notation of Poisson and Sasaki\cite{PoiSas}
for the source term and use $T_{\omega\ell m}$.

Note that Chrzanowski\cite{Chrz} uses different notation for
the source term $T$ in the Teukolsky equation.
Denoting Chrzanowski's $T$ by $T_{\rm C}$, it is related to
$T_{\omega\ell m}$ and ${}_{-2}{\cal T}_{\omega\ell m}$ as
\begin{eqnarray}
T_{\rm C} =- 2\,\pi \, T_{\omega \ell m}
=-4\,\pi\,r^2\,{}_{-2}{\cal T}_{\omega\ell m} \,.
\end{eqnarray}


\end{document}